\documentclass{revtex4}
\usepackage[T1]{fontenc}
\usepackage[utf8]{inputenc}
\usepackage{amsmath}
\usepackage{amsfonts}
\usepackage{emptypage}
\usepackage{mathrsfs}
 \usepackage{tikz}
\usepackage{graphicx}
\usepackage{wrapfig}
\usepackage[bf, small]{caption}
\usepackage{empheq}
\usepackage{subfigure}
\usepackage{graphics}
\usepackage{algorithm}
\usepackage{amssymb}
\usepackage{amsfonts}
\usepackage{latexsym}
\usepackage{listings}
\usepackage{calrsfs}
\usepackage{graphicx} 

\begin{document}


\title{Noise seeded oscillators: on the role of demographic fluctuations in a multi-populations model}

\author{Francesca Di Patti}

\affiliation{Dipartimento di Matematica e Informatica ``Ulisse Dini'', Università degli Studi di Firenze, viale Morgagni 67/a, 50134 Firenze, Italy.}

\altaffiliation{INFN, Sezione di Firenze, via G. Sansone 1, 50019 Sesto Fiorentino, Italy.}
\altaffiliation{CSDC - Centro Interdipartimentale per lo Studio delle Dinamiche Complesse, via G. Sansone 1, 50019 Sesto Fiorentino, Italy.}

\author{Duccio Fanelli}

\affiliation{Dipartimento di Fisica e Astronomia, Università degli Studi di Firenze, via G. Sansone 1, 50019 Sesto Fiorentino, Italy.}

\altaffiliation{INFN, Sezione di Firenze, via G. Sansone 1, 50019 Sesto Fiorentino, Italy.}
\altaffiliation{CSDC - Centro Interdipartimentale per lo Studio delle Dinamiche Complesse, via G. Sansone 1, 50019 Sesto Fiorentino, Italy.}

\author{Perla Rosi}
\affiliation{Dipartimento di Matematica e Informatica ``Ulisse Dini'', Università degli Studi di Firenze, viale Morgagni 67/a, 50134 Firenze, Italy.}

\email{francesca.dipatti@unifi.it}


\begin{abstract}
Stochastic oscillations can emerge from a two-population model as triggered by endogenous finite size fluctuations. Here, an extended dynamical scenario is considered in which a third fluctuating species is added to a proto-typical scheme of neuronal interaction. As we shall prove both analytically and numerically, the third added species can enhance or even suppress the emergence of quasi-cycles, namely the coherent oscillations of the two original populations, as instigated by the demographic noise component. In general, investigating the coupled dynamics of noisy oscillators of the type considered could yield an extended framework for synchronization studies, beyond the pioneering setting introduced by Kuramoto.
\end{abstract}

\maketitle

\section{Introduction}
The study of synchronization phenomena has become a central topic in the analysis of complex systems, as coordinated behavior emerges across a broad range of natural and artificial systems, including neural populations, biological rhythms, power grids, and social interactions \cite{Pikovsky_Rosenblum_Kurths_2001, Strogatz2003}. Among the theoretical frameworks developed to understand collective synchronization, the Kuramoto model has played a pivotal role \cite{Kuramoto75}. Introduced as a minimal yet analytically tractable description of coupled oscillators, the Kuramoto model captures how a population of heterogeneous units can spontaneously move from incoherent dynamics to collective synchronization through interactions. Its simplicity and versatility have made it a paradigmatic model for investigating synchronization transitions, collective behaviors, and self-organized phenomena in complex systems \cite{Acebron, Kuramoto84, Stogatz2000}. The Kuramoto model, and its main alternatives, focus on phase synchronization. This latter often assumes amplitude to be constant or irrelevant (as e.g. in the classic Kuramoto model). But in real oscillators, amplitude dynamics matter immensely. True oscillators have amplitude stability, which phase synchronization models often gloss over.
As a relevant example, in biological contexts, and specifically in neuroscience, collective dynamics arise from interactions between populations of homologous entities whose activity is governed not only by phase relationships but it is primarily instigated by the nonlinear excitation and inhibition loops \cite{Kandel}. In this context, neural mass models provide a more suitable mesoscopic description of individual oscillators, and their emerging orchestrated dynamics. Among different dynamical schemes in computational neuroscience, the Wilson–Cowan model stands as a cornerstone framework to illustrate the temporal evolution of interacting excitatory and inhibitory neuronal populations \cite{Wilson72, Wilson73}. By incorporating nonlinear population responses and feedback mechanisms, the Wilson-Cowan model offers a rich description of collective neural activity and bridges microscopic interactions with macroscopic emergent dynamics.
As such, it has become the reference standard in the theoretical description of neural population dynamics, providing insight into a broad spectrum of collective neural phenomena, ranging from visual hallucinations and binocular rivalry to pathological states such as epilepsy \cite{Wilson2021}. Owing to its versatility and analytical tractability, it is now regarded as one of the most influential mean-field models in theoretical neuroscience.  Despite its success, the original formulation of the model is entirely deterministic and therefore neglects the effects of stochasticity. Accounting for fluctuations is however mandatory from a wide biological perspective: noise can in fact play a fundamental dynamical role by triggering collective response rather than representing a mere source of undesired perturbation \cite{DiPatti2026}. In particular, intrinsic fluctuations may actively induce self-organized behaviors even in parameter regions where the deterministic dynamics predicts a stable fixed point \cite{predatorPreyMcKane, Biancalani2010, DiPatti2018, DiPatti2023}. Motivated by these observations, stochastic reformulations of the Wilson–Cowan framework have been proposed \cite{Cowan2016}. Among these, recent studies have shown that endogenous fluctuations originating from the finite size of the system can seed synchronized quasi-cycles in excitatory populations distributed across network nodes \cite{Zancok2017, Zagli2017}. These findings emphasize the constructive role that demographic noise may play in shaping collective neural dynamics. This contribution is indeed positioned in this setting and aims at further elucidating the emergence of complex, seemingly regular, oscillations as sustained by the intrinsic finite size noise. Beyond the case here explored, the coupled dynamics of the noise triggered oscillators could be considered as a generalized framework of possible exploration, beyond  phase oscillator models inspired to the seminal work of Kuramoto.

A limitation of the standard stochastic formulation is that it generally relies on the assumption of moderate or low copy numbers for the interacting species. While this assumption is natural in many biochemical contexts, its applicability to neuronal populations may be less immediate and can therefore raise conceptual and modeling concerns. Nevertheless, neurons or, more generally, interacting biological units, often communicate through intermediary chemical species whose concentrations may indeed be sufficiently small for stochastic effects to become relevant, even when the neuronal populations themselves are large \cite{Pahle2009,Lecca2017}. In such situations, stochasticity is essential only for a subset of the dynamical variables, whereas the remaining degrees of freedom may still be accurately described within a deterministic framework \cite{Bressloff2018,Wylie2006}.

The aim of the present work is therefore to investigate hybrid dynamical scenarios in which stochastic  variables with different characteristic sizes coexist and interact. In particular, we seek to determine whether fluctuations acting only on a subset of the system variables are sufficient to sustain or suppress the emergence of quasi-cycles, the coherent oscillation stimulated by the endogenous noise component.
\section{The three-species stochastic model}
We consider an extension of the single-node model originally introduced in \cite{Zancok2017} in the context of neuroscience. The model describes the dynamics of two interacting neuronal populations composed of individuals of type $X$ and $Y$, where $X$ denotes excitatory neurons and $Y$ inhibitory ones. In the present work, we extend the original framework by introducing  additional interactions mediated by a third species, denoted by $Z$. We assume that the populations $X$ and $Y$ are characterized by a volume $V$, while the species $Z$ has a distinct volume $V_1$. The aforementioned volumes are a proxy of the finite size nature of the involved populations. As we shall see, when both $V$ and $V_1$ go to infinity, the intrinsic noise gets suppressed and the system converges to its deterministic limit.

The stochastic dynamics of the $X$ and $Y$ populations is governed by the same set of reactions as in \cite{Zancok2017}:
\begin{equation}\label{eq:chem1}
\begin{aligned}
X &\xrightarrow[]{1} \varnothing \\
Y &\xrightarrow[]{1} \varnothing \\
\varnothing &\xrightarrow[]{f(s_x)} X \\
\varnothing &\xrightarrow[]{f(s_y)} Y 
\end{aligned}
\end{equation}
where $f(s)$ is a sigmoid function which can be cast in the form:
\begin{equation} \label{eq:sigmoid}
    f(s)=\displaystyle{\frac{1}{1+be^{-s}}}
\end{equation}
whose steepness is controlled by the parameter $b$. The arguments of the sigmoid are given by
\begin{equation*}
\begin{split}
    s_x= -r\biggl(\frac{n_y}{V}-\frac{1}{2}\biggr) \\
    s_y= r\biggl(\frac{n_x}{V}-\frac{1}{2}\biggr)
    \end{split}
\end{equation*}
with $r$ a positive free parameter and $n_x$, $n_y$ denoting the instantaneous numbers of $X$ and $Y$ individuals, respectively. We then introduce the reactions describing the dynamics of the additional species $Z$ and its interaction with $X$ and $Y$:
\begin{equation}\label{eq:chem2}
\begin{aligned}
Z &\xrightarrow[]{\delta_z} \varnothing, \\
\varnothing &\xrightarrow[]{\alpha_z} Z \\
X + Z &\xrightarrow[]{\gamma} Z \\
Y + Z &\xrightarrow[]{\gamma} Z.
\end{aligned}
\end{equation}

The stochastic dynamics of the state vector $\mathbf{n}(t)=(n_x(t), n_y(t), n_z(t))$ is ruled by the master equation, the differential equation for the probability $P(\mathbf{n}, t)$ of observing the system in the state $\mathbf{n}$ at time $t$. Its generic form reads
\begin{equation} \label{eq:ME_generic}
    \frac{d P}{d t}(\mathbf{n},t)= \sum_{\mathbf{n}\neq \mathbf{n'}}T(\mathbf{n}|\mathbf{n'})P(\mathbf{n'},t)-T(\mathbf{n'}|\mathbf{n})P(\mathbf{n},t)
\end{equation}
where $T(\mathbf{n}|\mathbf{n'})$ denotes the probability rate that the system undergoes the transition $\mathbf{n'}\to\mathbf{n}$. The explicit form of the transition rates are listed in Eq. \eqref{eq:transitions} of section \ref{sec:KM}. Multiplying both sides of Eq. \eqref{eq:ME_generic} by, respectively $n_x$, $n_y$ and $n_z$, and summing over $\mathbf{n}$ one can derive the equations governing the average quantities $<n_l>=\sum_\mathbf{n} n_l P(\mathbf{n}, t)$ for $l=x,y,z$:
\begin{equation}
    \begin{cases}
        \displaystyle{\frac{d}{dt}<n_x>=<f(s_x)>-\biggl<\frac{n_x}{V}\biggr>-\gamma\biggl<\frac{n_x}{V}\frac{n_z}{V_1}\biggr>} \\
        \\
        \displaystyle{\frac{d}{dt}<n_y>=<f(s_y)>-\biggl<\frac{n_y}{V}\biggr>-\gamma\biggl<\frac{n_y}{V}\frac{n_z}{V_1}\biggr>} \\
        \\
        \displaystyle{\frac{d}{dt}<n_z>=\alpha_z -\delta_z\biggl<\frac{n_z}{V_1}\biggr>}.
    \end{cases}
\end{equation}
As anticipated above, when both volumes, $V$ and $V_1$, become large, while keeping the ratio $V_1/V$ constant,  the previous system reduces to its deterministic mean-field version
\begin{equation}\label{eq:MF3}
    \begin{cases}
        \displaystyle{\frac{dx}{d\tau_1}=\biggl[f\biggl(-r\biggl(y-\frac{1}{2}\biggr)\biggr)-x-\gamma xz\biggr]\frac{V_1}{V} }\\
        \\
        \displaystyle{\frac{dy}{d\tau_1}=\biggl[f\biggl(r\biggl(x-\frac{1}{2}\biggr)\biggr)-y-\gamma yz\biggr]\frac{V_1}{V}} \\
        \\
        \displaystyle{\frac{dz}{d\tau_1}=\alpha_z-\delta_z \: z}
    \end{cases}
\end{equation}
for the deterministic concentrations $x$, $y$ and $z$, defined as 
\begin{equation}
    x=\lim_{V\to +\infty} \biggl<\frac{n_x}{V}\biggr>, \quad y=\lim_{V\to +\infty}\biggl<\frac{n_y}{V}\biggr>, \quad z=\lim_{V_1\to +\infty} \biggl<\frac{n_z}{V_1}\biggl>.
\end{equation}
The appearance of the ratio $V_1/V$ in the first two equations of system \eqref{eq:MF3} originates from the introduction of the rescaled time variable $\tau_1 = t / V_1$. From the system of equations \eqref{eq:MF3}, it follows immediately that the variable $z$ admits a steady state given by $z^*=\dfrac{\alpha_z}{\delta_z}$ while the remaining variables display equilibrium solutions specified by conditions $x^*=y^*=\dfrac{1}{2}$, as in the original model, provided the parameter $b$ of the sigmoidal function \eqref{eq:sigmoid} satisfies  $b=\dfrac{1-\gamma z^*}{1+\gamma z^*}$. This is the setting that we shall adopt in the following. Since the parameter $b$ must be positive, this condition is fulfilled only if the following constraint holds:
\begin{equation}\label{eq:condition_b}
    1-\gamma\frac{\alpha_z}{\delta_z}>0 \: .
\end{equation}
The stability of the aforementioned equilibrium point can be assessed by computing the eigenvalues of the Jacobian matrix associated with system \eqref{eq:MF3}:
\begin{equation}\label{eq:JfixedPoint}
    \mathcal{J}(x^*, y^*, z^*)= \frac{V_1}{V}
    \begin{pmatrix}
        \displaystyle{-1-\gamma\frac{\alpha_z}{\delta_z} }& \displaystyle{-\frac{r}{4}\biggl(1-\gamma^2\frac{\alpha_z^2}{\delta_z^2}\biggr)} &\displaystyle{ -\frac{\gamma}{2}} \\
        & & \\
        
        \displaystyle{\frac{r}{4}\biggl(1-\gamma^2\frac{\alpha_z^2}{\delta_z^2}\biggr)} & \displaystyle{-1-\gamma\frac{\alpha_z}{\delta_z}} & \displaystyle{-\frac{\gamma}{2} }\\
        & & \\
        0 & 0 & \displaystyle{-\delta_z\frac{V}{V_1}}
    \end{pmatrix}    
\end{equation}
The above Jacobian matrix admits a real negative eigenvalue  $\lambda_1= -\delta_z$,   and a pair of complex conjugate eigenvalues $ \lambda_{2,3}=\dfrac{V_1}{V}\biggl[-1-\gamma\dfrac{\alpha_z}{\delta_z}\pm i\frac{r}{4}\biggl(1-\gamma^2\dfrac{\alpha_z^2}{\delta_z^2}\biggr)\biggr]$. Since the real parts of all eigenvalues are negative, the equilibrium point is asymptotically stable. Importantly, however, the presence of a non-zero imaginary part opens the possibility for stochastic oscillations around the fixed point. Although the deterministic dynamics exhibits damped oscillations, intrinsic noise can sustain these fluctuations, giving rise to so-called quasi-cycles. 

\begin{figure}[tb]
    \centering
    \includegraphics[scale=0.2]{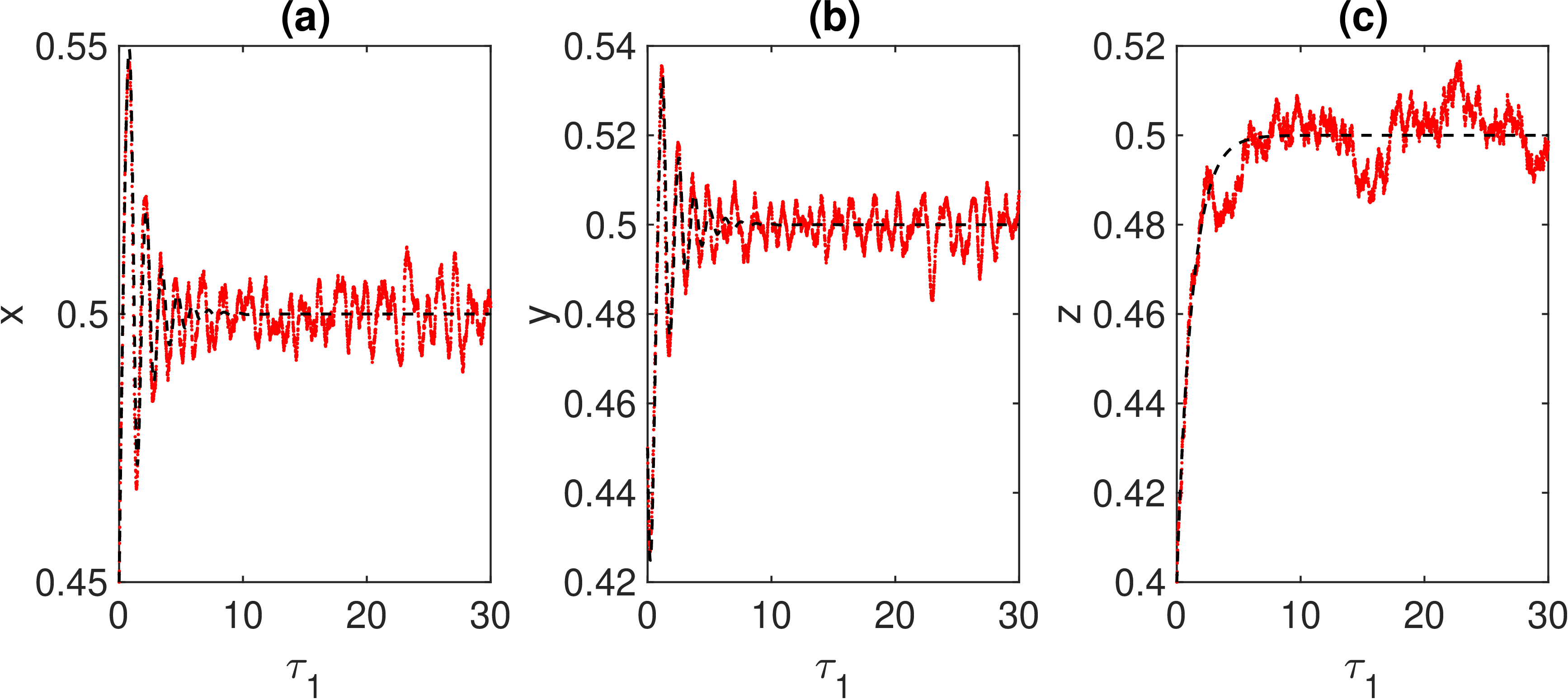}
    \caption{Comparison between the numerical integration of Eq. \eqref{eq:MF3} (dashed black line) and Eq. \eqref{eq:langevinKM} (red points). The parameters of the model are set as    $r=50$, $\gamma=0.9$, $\alpha_z=0.4$, $\delta_z=0.8$, $V=200$ and $V_1=100$. The differential equations are integrated numerically starting from the initial condition $[0.45 \: \:  0.45 \: \:  0.4]$ using a time step  $\delta_{\tau_1}=0.025$. }
    \label{fig:trajectoriesLang}
\end{figure}

To investigate this possibility, we apply the Kramers–Moyal expansion to the stochastic system. Following the derivation detailed in Section \ref{sec:KM}, we obtain a set of nonlinear Langevin equations:
\begin{equation}\label{eq:langevinKM}
 \begin{cases}
     \displaystyle{\frac{dx}{d\tau_1}=\frac{V_1}{V}(f(s_x)-x-\gamma xz)+\frac{1}{\sqrt{V_1}}\frac{V_1}{V}\sqrt{x+\gamma xz+f(s_x)} \ \lambda^{(1)}}  \\
     \\
     \displaystyle{\frac{dy}{d\tau_1}= \frac{V_1}{V}(f(s_y)-y-\gamma yz)+\frac{1}{\sqrt{V_1}}\frac{V_1}{V}\sqrt{y+\gamma yz+f(s_y)} \ \lambda^{(2)}} \\
     \\
     \displaystyle{\frac{dz}{d\tau_1}=\alpha_z-z\delta_z+\frac{1}{\sqrt{V_1}}\sqrt{z\delta_z+\alpha_z} \ \lambda^{(3)}} 
     \end{cases}  
 \end{equation}
Numerical simulations  of this stochastic system indeed display sustained stochastic oscillations, confirming the presence of noise-induced quasi-cycles for species $X$ and $Y$ (see Fig. \ref{fig:trajectoriesLang}).

\section{Role of  species $Z$ in preventing oscillations}
We now want to investigate how the emergent dynamics is affected by the intrinsic noise component as associated to the third species. This latter acts as a sort of external (stochastic) mediator which can self-consistently modulate the 
inherent dynamics of the paired neuronal populations $X$ and $Y$. For the above purpose we put forward the following ansatz
\begin{equation*}
\begin{aligned}
     x & =\biggl(\lim_{V\to+\infty}\frac{n_x}{V}\biggr)+\frac{\xi_x}{\sqrt{V_1}}=\frac{1}{2}+\frac{\xi_x}{\sqrt{V_1}} \\
    y & =\biggl(\lim_{V\to+\infty}\frac{n_y}{V}\biggr)+\frac{\xi_y}{\sqrt{V_1}}=\frac{1}{2}+\frac{\xi_y}{\sqrt {V_1}} \\
    z & =\biggl(\lim_{V_1\to+\infty}\frac{n_z}{V_1}\biggr)+\frac{\xi_z}{\sqrt{V_1}}=\frac{\alpha_z}{\delta_z}+\frac{\xi_z}{\sqrt{V_1}}
\end{aligned}  
\end{equation*}
and expand the system in terms of $1/\sqrt{V_1}$. Following the detailed derivation reported in section \ref{sec:LN}, we end up with a system of coupled linear Langevin equations for the vector of fluctuations $\boldsymbol{\zeta}=(\xi_x, \xi_y, \xi_z)$, namely:
\begin{equation}\label{eq:Lang_lin}
    \frac{d\zeta_i}{d\tau_1}=(\mathcal{J}\zeta)_i+G_i\lambda^{(i)}
\end{equation}
where the matrix $G$ reads
\begin{equation}\label{eq:matrixG}
    G=
    \begin{pmatrix}
        \displaystyle{\frac{V_1}{V}\sqrt{1+\gamma\frac{\alpha_z}{\delta_z}}} & 0 & 0 \\
        & & \\
        0 & \displaystyle{\frac{V_1}{V}\sqrt{1+\gamma\frac{\alpha_z}{\delta_z}} }& 0 \\
        & & \\
        0 & 0 & \sqrt{2\alpha_z}
    \end{pmatrix}
\end{equation}
while $\mathcal{J}$ is the matrix defined by Eq. \eqref{eq:JfixedPoint}. The term $\lambda^{(i)}$ denotes the $i-th$ component of the three-dimensional vector $\boldsymbol{\lambda}$ whose elements are Gaussian white noise variables with unit variance. 

Equation \eqref{eq:Lang_lin} can be solved in  Fourier space. Applying the temporal Fourier transform we get
 \begin{equation}\label{eq:LangFourier}
     -i\omega_{\tau_1}\tilde{\zeta}_i(\omega_{\tau_1})=(\mathcal{J}\tilde{\zeta})_i+G_i\tilde{\lambda}^{(i)},
 \end{equation}
 and, defining the matrix $\Phi=-\mathcal{J}-i\omega_{\tau_1}\mathbb{I}_{3}$ where $\mathbb{I}_{3}$ is the $3 \times 3$ identity matrix and collecting the noise terms in the  vector $ \boldsymbol{\tilde{\zeta}}$, the solution of Eq. \eqref{eq:LangFourier} is
 \begin{equation}
     \boldsymbol{\tilde{\zeta}}(\omega_{\tau_1})=\Phi^{-1}G\boldsymbol{\tilde{\lambda}} \quad .
     \label{soluzione Langevin}
 \end{equation}
 With this solution we can evaluate the power spectrum density matrix whose entries are
 \begin{equation}\label{eq:powerSpectrumLN}
     P_{ij}(\omega_{\tau_1})=<\tilde{\zeta}_i(\omega_{\tau_1})\tilde{\zeta}_j^*(\omega_{\tau_1})>= \left ( \Phi^{-1} G G ^t (\Phi^{-1})^\dagger \right )_{ij}. 
 \end{equation}
 where the indices $i,j=1,2,3$ label the species, corresponding to $X$, $Y$ and $Z$, respectively. The explicit form of matrix  $\Phi$ is: 
 \begin{equation*}
     \Phi= 
     \begin{pmatrix}
         \displaystyle{\frac{V_1}{V}\biggl(1+\gamma\frac{\alpha_z}{\delta_z}\biggr)-i\omega_{\tau_1} }& \displaystyle{\frac{V_1}{V}\frac{r}{4}\biggl(1-\gamma^2\frac{\alpha_z^2}{\delta_z^2}\biggr)} & \displaystyle{\gamma\frac{ V_1}{2V}} \\
         & & \\
         \displaystyle{-\frac{V_1}{V}\frac{r}{4}\biggl(1-\gamma^2\frac{\alpha_z^2}{\delta_z^2}\biggr)} & \displaystyle{\frac{V_1}{V}\biggl(1+\gamma\frac{\alpha_z}{\delta_z}\biggr)-i\omega_{\tau_1}} &\displaystyle{\gamma \frac{ V_1}{2V}} \\
         & & \\
         0 & 0 & \displaystyle{\delta_z-i\omega_{\tau_1}}
     \end{pmatrix}
 \end{equation*}
Following the argument introduced in \cite{predatorPreyMcKane}, the position  of the peak of the power spectrum can be inferred by minimizing the denominator of the power spectrum. In this specific case, the relevant quantity can be explicitly written as: 
\begin{equation}\label{eq:denominatorP}
     \det(\Phi) \det(\Phi^{\dagger}) = \left ( \delta_z^2 + \omega_{\tau_1}^2 \right ) 
    \left \{ 
    \left [ \omega_{\tau_1}^2 -A \right ]^2  + 2 \omega_{\tau_1}^2  B + C
         \right \}
\end{equation} 
with 
\begin{equation*}
   \begin{aligned}
    A &= \dfrac{r^2}{16} \dfrac{V_1^2}{V^2} \left ( 1 -\gamma^2 \dfrac{\alpha_z^2}{\delta_z^2}\right )^2 \\
    B &= \dfrac{V_1^2}{V^2} \left ( 1 +\gamma \dfrac{\alpha_z}{\delta_z}\right )^2 \\
    C & =\dfrac{V_1^4}{V^4} \left ( 1 +\gamma \dfrac{\alpha_z}{\delta_z}\right )^4 
    \left [  1+ \dfrac{r^2}{8} \left (  1 -\gamma \frac{\alpha_z}{\delta_z}\right )\right ]
 \end{aligned} 
\end{equation*}
 Expression \eqref{eq:denominatorP} is minimized for $\omega_{\tau_1}^2\approx A$, that is for 
\begin{equation}\label{eq:omegaPeak}
 \omega_{\tau_1}\approx \omega_{\tau_1}^* = \frac{r}{4}\dfrac{V_1}{V}\biggl(1-\gamma^2\dfrac{\alpha_z^2}{\delta_z^2}\biggr).
\end{equation}
which is positive defined in light of the the constraint imposed by Eq. \eqref{eq:condition_b}.

\begin{figure}[tb]
    \centering
    \includegraphics[scale=0.2]{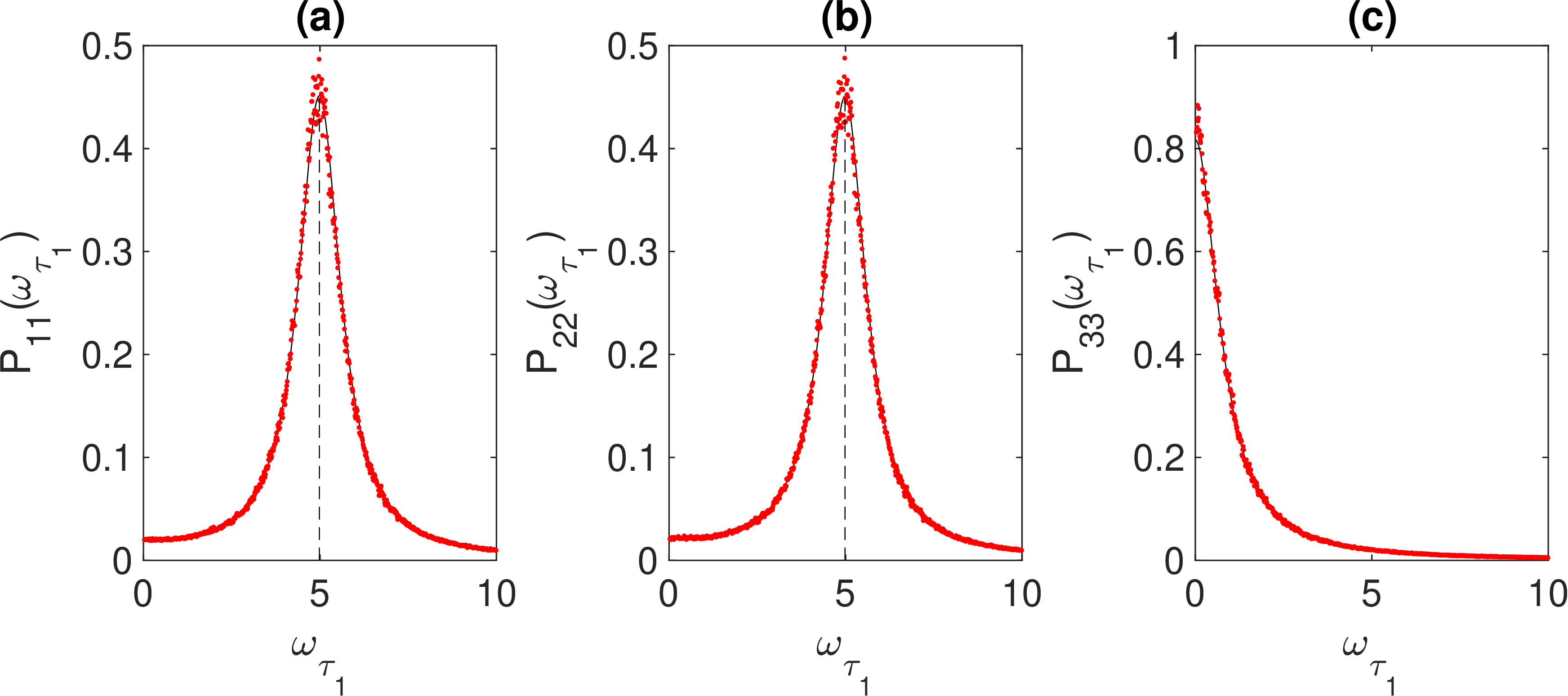}
    \caption{Comparison between the theoretical power spectrum (solid black line) obtained through the linear noise approximation (Eq. \eqref{eq:powerSpectrumLN}), and numerical simulations (red points), computed by averaging over $500$ independent realizations of the Langevin equation \eqref{eq:langevinKM}. Panel (a) refers to species $X$, (b) is relative to species $Y$ while panel (c) shows the power spectrum for species $Z$. The Langevin equation was integrated using the Euler–Maruyama algorithm \cite{EulerMaruyama} with time step $\delta_{\tau_1}=0.0017$ over the time interval $[0, 500]$, starting from the equilibrium point $(x^*, y^*, z^*)$ as the initial condition. The model parameters are the same as those used in Fig. \ref{fig:trajectoriesLang}. The vertical dashed lines present in panels (a) and (b) mark the position of the peak predicted by Eq. \eqref{eq:omegaPeak}.}
    \label{fig:Ps_vs_Lang}
\end{figure}

Figure \ref{fig:Ps_vs_Lang} shows the agreement between the power spectrum obtained through the linear noise approximation (Eq. \eqref{eq:powerSpectrumLN}) and the stochastic simulations. It is also evident that the estimate provided by Eq. \eqref{eq:omegaPeak} accurately predicts the position of the spectral peak (see the dashed vertical lines as reported in panels (a) and (b) of Fig. \ref{fig:Ps_vs_Lang}). Notably, and as expected, the power spectrum for species $Z$ does not exhibit any peak, indicating that the noisy signal observed in Fig. \ref{fig:trajectoriesLang}c consists of simple fluctuations, superposed to the deterministic mean-field trajectory, rather than quasi-cycles.
 
\begin{figure}[tb]
    \centering
    \includegraphics[scale=0.2]{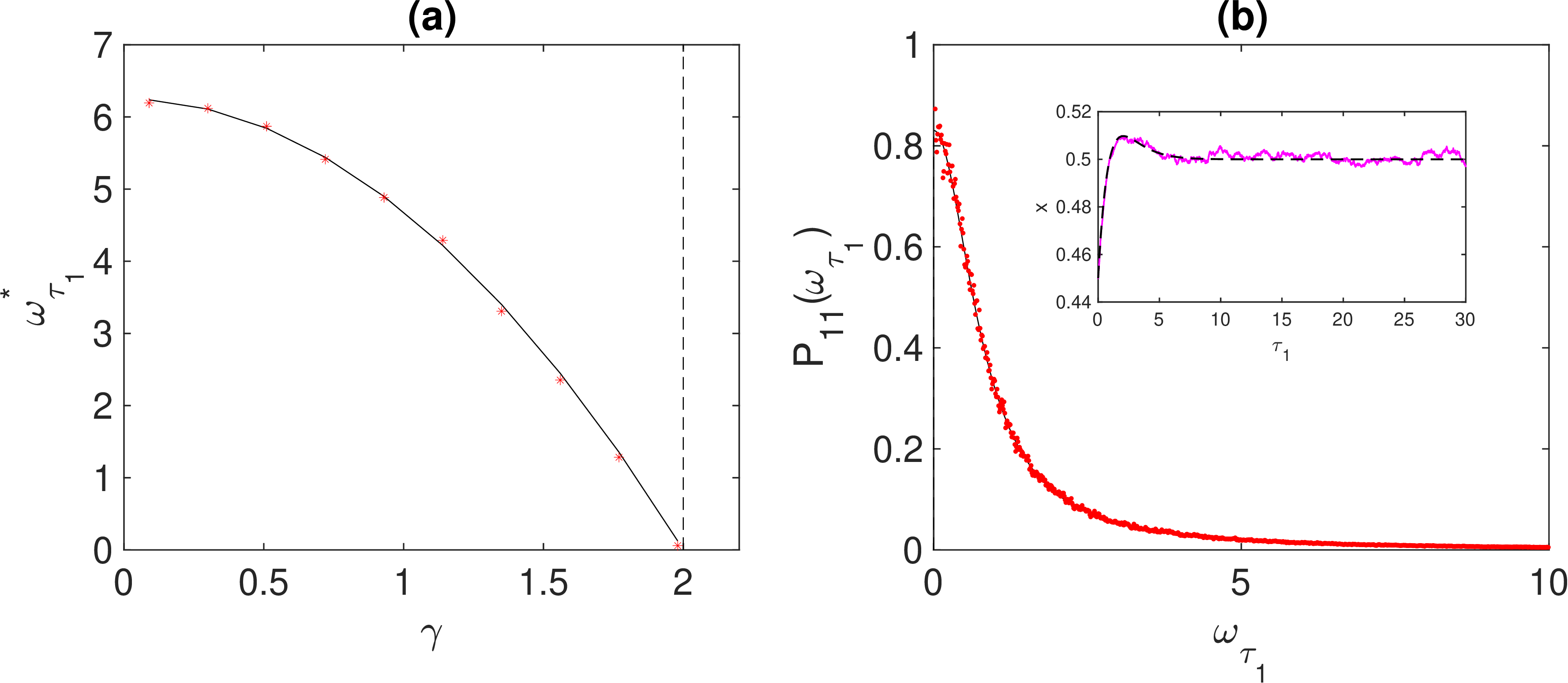} 
    \caption{Panel (a): Position of the peak of the power spectrum for species $X$ as a function of $\gamma$. The solid line denotes the theoretical prediction given by Eq. \eqref{eq:omegaPeak} whereas the red dots represent the position of the peak obtained averaging $300$ numerical integration of the Langevin equation \eqref{eq:langevinKM} through the Euler-Maruyama method \cite{EulerMaruyama} with time step $\delta \tau_1 =0.0013$ over the time interval $[0,  500]$. Panel (b): Power spectrum of fluctuation relative to species $X$ for $\gamma$ close to the critical value $\delta_z/\alpha_z$ that is $\gamma = \delta_z/\alpha_z - \varepsilon$ with $\varepsilon = 10^{-6}$. The inset shows a typical stochastic trajectory for species $X$ obtained by numerical integration of the Langevin equation \eqref{eq:langevinKM} (noisy magenta line) and the deterministic equation \eqref{eq:MF3} (dashed black line) starting from the initial configuration $[0.45, 0.45, 0.4]$.  Unspecified parameters are the same as those used in Fig. \ref{fig:trajectoriesLang}.}
    \label{fig:peak_vs_gamma}
\end{figure}

Let us now focus on the prediction given by Eq. \eqref{eq:omegaPeak}. 
The position of the peak of the power spectrum, as expected, is modulated both by the ratio between the characteristic sizes of the two families of populations and by the kinetic parameters of species $Z$. In particular, it is interesting to analyze the dependence of the peak frequency 
$\omega_{\tau_1}$ on the parameter $\gamma$, which mediates the interaction between species $Z$ and the other two. As can be appreciate by inspection of Fig. \ref{fig:peak_vs_gamma}a, the peak position reaches its maximum values as  $\gamma$ approaches $0$ (corresponding to the case of two uncoupled families of species), whereas increasing $\gamma$ up to $\dfrac{\delta_z}{\alpha_z}$, the largest possible value as allowed by condition (\ref{eq:condition_b}), the peak position tends to zero. 
From this last observation it is therefore clear that, in the limit
\begin{equation*}
    \gamma\frac{\alpha_z}{\delta_z}\to 1
\end{equation*}
the effect of species $Z$ on the other two is to inhibit the emergence of quasi-cycles. In this limit, in fact, the power spectra of species $X$ and $Y$ tend to attain their maximum at  $\omega_{\tau_1}=0$, thus losing the periodicity of the generated signal. In Fig. \ref{fig:peak_vs_gamma}b an example of the power spectrum in this limit is shown. In the annexed inset, the corresponding stochastic and deterministic time series for species $X$ is reported: quasi-cycles are no longer present, as correctly anticipated. Identical considerations naturally apply to species $Y$. 
 \section{Noise amplification in large neuronal populations}
 In this section we investigate the limiting case $V >> V_1$, corresponding to a regime in which the neuronal population is much larger than the characteristic size associated with species $Z$. In this formal limit, we can perform a consistent elimination of the stochastic density referred to species $Z$ to yield a compact system of two coupled stochastic differential equations for the remaining species. We begin by rewriting  Eq. \eqref{eq:langevinKM} in the original (unscaled) time $t$ yielding:
 \begin{equation*}
 \begin{cases}
     \displaystyle{\frac{dx}{dt}=\frac{1}{V}(f(s_x)-x-\gamma xz)+\dfrac{1}{\sqrt{V_1}}\frac{1}{V}\sqrt{x+\gamma xz+f(s_x)} \ \lambda^{(1)}}  \\
     \\
     \displaystyle{\frac{dy}{dt}= \dfrac{1}{V}(f(s_y)-y-\gamma yz)+\dfrac{1}{\sqrt{V_1}}\frac{1}{V}\sqrt{y+\gamma yz+f(s_y)} \ \lambda^{(2)}} \\
     \\
     \displaystyle{\dfrac{dz}{dt }= \dfrac{1}{V_1}\alpha_z-z\delta_z+\dfrac{1}{\sqrt{V_1}} \dfrac{1}{V_1}\sqrt{z\delta_z+\alpha_z} \ \lambda^{(3)}} 
     \end{cases}  \quad .
 \end{equation*}
 Performing the change of variables $\phi_x= \lim_{V \rightarrow \infty} x$ and $\phi_y= \lim_{V \rightarrow \infty} y$ and introducing the re-scaled time $\tau=t/V$, the previous set of equations becomes
  \begin{equation}\label{eq:langevinKM_tau1_tau2}
 \begin{cases}
     \dfrac{d \phi_x }{d \tau}=f(s_{\phi_x})-\phi_x-\gamma \phi_x z +\dfrac{1}{\sqrt{V_1}}\sqrt{\phi_x+\gamma  \phi_x z+f(s_{\phi_x})} \ \lambda^{(1)}  \\
     \\
    \dfrac{d \phi_y}{d \tau}= f(s_{\phi_y})-\phi_y-\gamma  \phi_y z)+\dfrac{1}{\sqrt{V_1}}\sqrt{ \phi_y+\gamma  \phi_y z+f(s_{\phi_y})} \ \lambda^{(2)} \\
     \\
     \dfrac{dz}{d \tau_1 }= \alpha_z-z\delta_z+\dfrac{1}{\sqrt{V_1}} \sqrt{z\delta_z+\alpha_z} \ \lambda^{(3)} 
     \end{cases}  \quad .
 \end{equation}
Notice that two different time variables appear, $\tau$ and $\tau_1=t/V_1$ defined  as in the previous section. This does not pose a problem for the subsequent analysis.  We now   expand $z$ around its steady state as $z= \bar{z} +\dfrac{1}{\sqrt{V_1}} \xi$ where $\xi$ is a stochastic variable. Substituting this expression into the equation for $z$ in system \eqref{eq:langevinKM_tau1_tau2} we obtain
 \begin{equation}
   \dfrac{1}{\sqrt{V_1}} \dfrac{d \xi}{d \tau_1} =  \alpha_z - \delta_z  \left ( \bar{z}  +\dfrac{1}{\sqrt{V_1}} \xi \right ) + \dfrac{1}{\sqrt{V_1}} \sqrt{ \left ( \bar{z}  +\dfrac{1}{\sqrt{V_1}} \xi  \right ) \delta_z+\alpha_z} \: \lambda^{(3)} \quad . 
 \end{equation}
 Expanding and neglecting higher-order terms, we match contributions at different orders  in $V_1^{-1/2}$. At order  $O(1)$ we recover  $\bar{z}=\alpha_z/\delta_z \equiv z^* $, while at order $1/\sqrt{V_1}$ we find the stochastic differential equation for $\xi$
 \begin{equation*}
      \dfrac{d \xi}{d \tau_1} = - \delta_z \xi + \sqrt{2 \alpha_z} \: \lambda^{(3)}
 \end{equation*}
 whose stationary solution is $\xi= \sqrt{\dfrac{\alpha_z}{\delta_z}} \eta$ where $\eta$ is a random variable from the normal distribution. The above result follows readily from the stationary solution of the Fokker-Planck equation associated to
the above linear Langevin equation. Hence: 
 \begin{equation}\label{eq:z_adia}
     z = \dfrac{\alpha_z}{\delta_z} + \dfrac{1}{\sqrt{V_1}} \sqrt{ \dfrac{ \alpha_z}{\delta_z }} \: \eta \: .
 \end{equation}
Substituting Eq.~\eqref{eq:z_adia} into the first two equations of system \eqref{eq:langevinKM_tau1_tau2}, we obtain
 \begin{equation} \label{eq:adiabaticEq}
     \begin{cases}
     \dfrac{d \phi_x }{d \tau}=f(s_{\phi_x})-\phi_x \left ( 1 + \gamma \dfrac{\alpha_z}{\delta_z} \right )
     +\dfrac{1}{\sqrt{V_1}}\sqrt{\phi_x+ \gamma \dfrac{\alpha_z}{\delta_z}  \phi_x  +  \gamma ^2\dfrac{\alpha_z}{\delta_z}  \phi_x^2 +f(s_{\phi_x})} \:  \eta  \\
     \\
     \dfrac{d \phi_x }{d \tau}=f(s_{\phi_y})-\phi_y \left ( 1 + \gamma \dfrac{\alpha_z}{\delta_z} \right )
     +\dfrac{1}{\sqrt{V_1}}\sqrt{\phi_y+ \gamma \dfrac{\alpha_z}{\delta_z}  \phi_y  +  \gamma ^2\dfrac{\alpha_z}{\delta_z}  \phi_y^2 +f(s_{\phi_y})} \:  \eta  \\
     \end{cases}
 \end{equation}
 We now assume that $\phi_x$ and $\phi_y$ fluctuate around their steady-state values $(x^*,y^*)=(1/2, 1/2)$ and write them as 
\begin{equation}
\begin{aligned}
    \phi_x &= \dfrac{1}{2} + \dfrac{1}{\sqrt{V_1}} \xi_x \\
    \phi_y &= \dfrac{1}{2} + \dfrac{1}{\sqrt{V_1}} \xi_y
\end{aligned}
\end{equation}
and perform an expansion of the non linear terms in Eq. \eqref{eq:adiabaticEq} to leading order in $1/\sqrt{V_1}$ obtaining
\begin{equation}\label{eq:approx_sigma}
\begin{aligned}
    f(s_{\phi_x}) \approx &  \dfrac{1}{1+b} -r  \dfrac{1}{\sqrt{V_1}}\dfrac{b}{\left ( 1 +b \right ) ^2} \xi_y \\
    f(s_{\phi_y}) \approx & \dfrac{1}{1+b} +r  \dfrac{1}{\sqrt{V_1}}\dfrac{b}{\left ( 1 +b \right ) ^2} \xi_x 
\end{aligned} \quad .
\end{equation}
Substituting Eq. \eqref{eq:approx_sigma} into Eq. \eqref{eq:adiabaticEq} and retaining only terms up to order $O(1)$ and $O(1/\sqrt{V_1})$ we end up with the following linear Langevin equation
\begin{equation}\label{eq:LangVinf}
    \dfrac{d}{d \tau} 
    \begin{pmatrix}
        \xi_x  \\ \xi_y
    \end{pmatrix}
    = J \begin{pmatrix}
        \xi_x \\ \xi_y
    \end{pmatrix} + \sqrt{1 + \gamma \dfrac{\alpha_z}{\delta_z} + \dfrac{\gamma^2}{4} \dfrac{\alpha_z}{\delta_z}}
    \begin{pmatrix}
        \eta \\  \eta
    \end{pmatrix}
\end{equation}
where $J$ is the $2 \times 2 $ sub-matrix obtained by removing  the third row and the third column from the Jacobian matrix $\mathcal{J}$ defined in Eq.  \eqref{eq:JfixedPoint}. Comparing \eqref{eq:LangVinf}
with Eqs. \eqref{eq:Lang_lin} and \eqref{eq:matrixG} we observe that the noise amplitude is increased by the positive term $\dfrac{\gamma^2}{4} \dfrac{\alpha_z}{\delta_z}$ thereby showing that, in the limit of large neuronal populations, the effect of species $Z$ is to enhance the strength of the (finite size) noise. 

\begin{figure}[tb]
    \centering
    \includegraphics[scale=0.6]{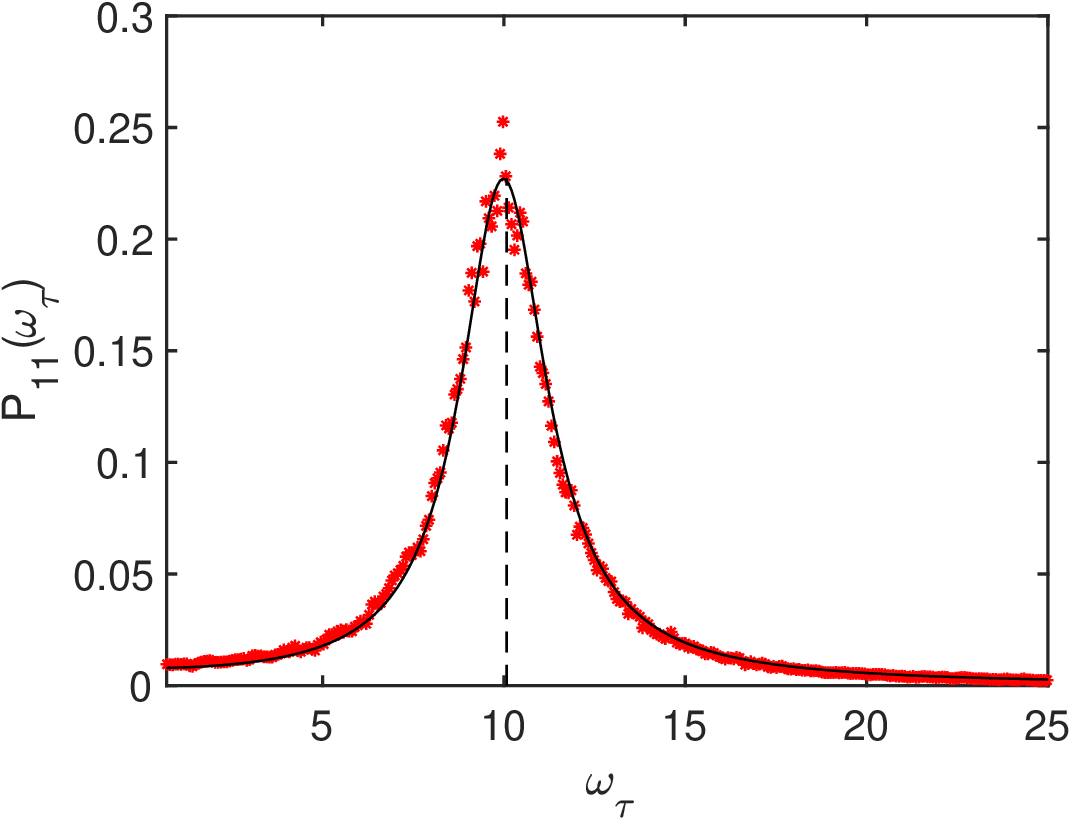}
    \caption{Power spectrum of fluctuations for species $X$. The solid black line corresponds to the theoretical prediction given by Eq. \eqref{eq:PS_2species} while red dots are obtained from numerical integration of the Langevin equation \eqref{eq:langevinKM} using the Euler-Maruyama algorithm \cite{EulerMaruyama} with time step $\delta \tau_1=0.0002$ over  $2 \cdot 10^7$ steps, and averaged over realizations $200$ realizations. The model parameters are $r=50$, $\gamma=0.9$, $\delta_z=0.8$, $\alpha_z=0.4$, $V_1=200$ and $V=10000$. To enable comparison between the two spectra, frequencies have been rescaled as  $\omega_{\tau}=V/V_1 \omega_{\tau_1}$. }
    \label{fig:psNoiseOnly_z}
\end{figure}

Also in this case, the power spectrum of the fluctuations can be computed and is found to have the same structure as Eq. \eqref{eq:powerSpectrumLN}:
\begin{equation}\label{eq:PS_2species}
     P_{ij}(\omega_{\tau_1})= \left [ 1 + \gamma \dfrac{\alpha_z}{\delta_z} + \dfrac{\gamma^2}{4} \dfrac{\alpha_z}{\delta_z} \right ] \left (\Phi^{-1} 
    \begin{pmatrix}
        1 & 1\\
        1 & 1
    \end{pmatrix}(\Phi^{-1})^\dagger \right )_{ij}. 
\end{equation}
where matrix $\Phi=-J-i\omega_{\tau}\mathbb{I}_{2}$ with $\mathbb{I}_{2}$ is the $2 \times 2$ identity matrix. Here indices $i$ and $j$ are limited to $\{1,2\}$. As before, the position of the spectral peak can be estimated by minimizing the denominator, yielding
\begin{equation}\label{eq:omegaPeak_onlyz}
\omega_{\tau}^*= \left ( 1 + \gamma \dfrac{\alpha_z}{\delta_z} \right ) \sqrt{1 +\dfrac{r^2}{16} \left (  1 - \gamma \dfrac{\alpha_z}{\delta_z} \right )^2 }
\end{equation}
The accuracy of the proposed approximation is confirmed by  Fig. \ref{fig:psNoiseOnly_z} which shows the comparison of the theoretical power spectrum \eqref{eq:PS_2species} with the analogous quantity obtained by processing the numerical trajectory. This latter is obtained by integrating the Langevin equation \eqref{eq:langevinKM} for the complete version of the model (i.e. by accounting for the simultaneous presence of the three species) operated in the relevant setting $V >> V_1$. 
\section{Conclusion}
We have here presented an extension of the model introduced in \cite{Zancok2017} by including an additional interacting species. We assumed that this new species is characterized by a different intrinsic size scale with respect to the two neuronal populations. This framework allowed us to investigate the role of species $Z$ in two distinct regimes. When the characteristic sizes of the two neuronal populations and species $Z$ are of the same order, the presence of $Z$ modulates the position of the oscillatory peak, up to a limiting case in which oscillations in the neuronal populations disappear altogether. Conversely, when the neuronal populations are assumed to be very large, the effect of the additional species is to enhance the intensity of the stochastic fluctuations, effectively amplifying the strength of the noise. This latter effect is captured analytically by resorting to a simplified two populations setting, which follows a self-consistent stochastic elimination of the third added species. In general, investigating the coupled dynamics of noisy oscillators of the type considered here could open novel avenues of exploration for the onset of synchronization, moving beyond the pioneering and still highly relevant work of Kuramoto.

\appendix
\section{The Kramers-Moyal expansion}\label{sec:KM}
The transition rates associated with the  stochastic rules \eqref{eq:chem1} and \eqref{eq:chem2} are given by
\begin{equation}\label{eq:transitions}
    \begin{aligned}
         T(n_x-1,n_y,n_z|\mathbf{n}) &=  \frac{n_x}{V}+\gamma\dfrac{n_x}{V}\dfrac{n_z}{V_1} \\
         T(n_x, n_y-1, n_z|\mathbf{n}) &=  \dfrac{n_y}{V}+\gamma\dfrac{n_y}{V}\dfrac{n_z}{V_1} \\
         T(n_x, n_y, n_z-1|\mathbf{n})& =\dfrac{n_z}{V_1}\delta_z\\
          T(n_x, n_y, n_z+1|\mathbf{n}) & = \alpha_z
    \end{aligned} \quad .
\end{equation}
The corresponding master equation reads 
\begin{equation} \label{eq:masterEquation} 
\begin{aligned}
         \frac{d P}{d t} (\mathbf{n}, t)&= T(\mathbf{n}|n_x+1, n_y, n_z)P(n_x+1, n_y, n_z, t)\\
        &+T(\mathbf{n}|n_x, n_y+1, n_z)P(n_x, n_y+1, n_z, t)\\
        &+T(\mathbf{n}|n_x, n_y, n_z+1)P(n_x, n_y, n_z+1, t)\\
        &+T(\mathbf{n}|n_x-1, n_y, n_z)P(n_x-1, n_y, n_z, t)\\
        &+T(\mathbf{n}|n_x, n_y-1, n_z)P(n_x, n_y-1, n_z, t)\\
        &+T(\mathbf{n}|n_x, n_1, n_z-1)P(n_x, n_y, n_z-1, t)\\
    &-T(n_x+1, n_y, n_z|\mathbf{n})P(\mathbf{n},t)
    -T(n_x, n_y+1, n_z|\mathbf{n})P(\mathbf{n},t)\\
    &-T(n_x, n_y, n_z+1|\mathbf{n})P(\mathbf{n},t)
    -T(n_x-1, n_y, n_z|\mathbf{n})P(\mathbf{n},t)\\
    &-T(n_x, n_y-1, n_z|\mathbf{n})P(\mathbf{n},t)-T(n_x, n_y, n_z-1|\mathbf{n})P(\mathbf{n},t)
\end{aligned} \quad .
\end{equation}
Introducing the step operators
\begin{equation*}
    \begin{split}
        \epsilon_x^\pm g(n_x, n_y, n_z)= g(n_x\pm 1, n_y, n_z), \\
        \epsilon_y^\pm g(n_x, n_y, n_z)= g(n_x, n_y\pm 1, n_z), \\
        \epsilon_z^\pm g(n_x, n_y, n_z)= g(n_x, n_y, n_z \pm 1).
    \end{split}
\end{equation*} 
where  $g(n_x, n_y, n_z)$ is a generic function, the master equation can be written in compact form as
\begin{multline} \label{eq:masterEquationStepOp} 
    \frac{d P}{d t}(\mathbf{n}, t)=[(\epsilon_x^+-1)T(n_x-1, n_y, n_z|\mathbf{n})+(\epsilon_x^--1)T(n_x+1, n_y, n_z|\mathbf{n})]P(\mathbf{n},t)\\
    +[(\epsilon_y^+-1)T(n_x, n_y-1, n_z|\mathbf{n})+(\epsilon_y^--1)T(n_x, n_y+1, n_z|\mathbf{n})]P(\mathbf{n},t)\\
    +[(\epsilon_z^+-1)T(n_x, n_y, n_z-1|\mathbf{n})+(\epsilon_z^--1)T(n_x, n_y, n_z+1|\mathbf{n})]P(\mathbf{n},t). 
\end{multline}
We now perform the Kramers-Moyal expansion. To this end we introduce the new variables 
$x=\frac{n_x}{V}$, $y=\frac{n_y}{V}$ and $z=\frac{n_z}{V_1}$ so that the step operators can be expressed as 
 \begin{align*}
     \epsilon ^\pm_x g(x,y,z)=g\biggl(x\pm\frac{1}{V},y,z\biggr) \\
     \epsilon^\pm_y g(x,y,z)=g\biggl(x,y\pm\frac{1}{V},z\biggr) \\
     \epsilon^\pm_z g(x,y,z)=g\biggl(x,y,z\pm \frac{1}{V_1}\biggr)
 \end{align*}
 and, in the limit of large volumes, we can  expand the step operators in powers of $1/\sqrt{V}$ and $1/\sqrt{V_1}$ as
 \begin{equation}
     \begin{aligned}\label{eq:expansionOperators}
     \epsilon_x^\pm \approx 1\pm \frac{1}{V}\frac{\partial}{\partial x}+\frac{1}{2V^2}\frac{\partial^2}{\partial x^2} \\
     \epsilon_y^\pm \approx 1\pm \frac{1}{V}\frac{\partial}{\partial y}+\frac{1}{2V^2}\frac{\partial^2}{\partial y^2} \\
     \epsilon_z^\pm \approx 1\pm \frac{1}{V_1}\frac{\partial}{\partial z}+\frac{1}{2V_1^2}\frac{\partial^2}{\partial z^2}.
 \end{aligned}
 \end{equation}
 
 Inserting  \eqref{eq:expansionOperators} into \eqref{eq:masterEquationStepOp} we get 
 \begin{align*}
     \frac{d P}{d  t}(x,y,z,t)= \frac{1}{V}\frac{\partial}{\partial x}[(T(n_x-1,n_y,n_z|\mathbf{n})-T(n_x+1,n_y,n_z|\mathbf{n}))P(\mathbf{n},t)]  \\
     +\frac{1}{V}\frac{\partial}{\partial y}[(T(n_x, n_y-1,n_z|\mathbf{n})-T(n_x, n_y+1, n_z|\mathbf{n}))P(\mathbf{n},y)]  \\
     + \frac{1}{V_1}\frac{\partial}{\partial z}[(T(n_x, n_y, n_z-1|\mathbf{n})-T(n_x, n_y, n_z+1|\mathbf{n}))P(\mathbf{n},t)] \\
     +\frac{1}{2V^2}\frac{\partial^2}{\partial x^2}[(T(n_x-1,n_y, n_z|\mathbf{n})+T(n_x+1, n_y, n_z|\mathbf{n}))P(\mathbf{n},t)] \\
     +\frac{1}{2V^2}\frac{\partial^2}{\partial y^2}[(T(n_x, n_y-1, n_z|\mathbf{n})+T(n_x, n_y+1, n_z|\mathbf{n}))P(\mathbf{n},t)] \\
     +\frac{1}{2V_1^2}\frac{\partial^2}{\partial z^2}[T(n_x, n_y, n_z-1|\mathbf{n})+T(n_x, n_y, n_z+1|\mathbf{n}))P(\mathbf{n}, t)]. 
 \end{align*}
 that, introducing the vector  $\mathbf{k}=(x,y,z)$ and rescaling time as  $\displaystyle{\tau_1=\frac{t}{V_1}}$, can be rewritten as the following Fokker-Planck equation 
 \begin{equation}\label{eq:Fokker-Planck}
     \frac{d P}{d \tau_1}(\mathbf{k},t)=-\sum_{i=1}^3\frac{\partial}{\partial k_i}A_iP(\mathbf{n},t)+\sum_{i=1}^3\frac{1}{2V_1}\frac{\partial^2}{\partial k_i^2}B_iP(\mathbf{n},t)
\end{equation}
with 
\begin{equation*}
\mathbf{A}= 
\begin{pmatrix}
\dfrac{V_1}{V}(T(n_x+1,n_y, n_z|\mathbf{n})-T(n_x-1, n_y, n_z|\mathbf{n})) \\
\\
\dfrac{V_1}{V}(T(n_x, n_y+1, n_z|\mathbf{n})-T(n_x, n_y-1, n_z|\mathbf{n})) \\
\\
T(n_x, n_y, n_z+1|\mathbf{n})-T(n_x, n_y, n_z-1,|\mathbf{n})
 \end{pmatrix}
\end{equation*}
and 
\begin{equation}\label{eq:vectorB}
\mathbf{B}=
 \begin{pmatrix}
\dfrac{V_1^2}{V^2}(T(n_x-1,n_y,n_z|\mathbf{n})+T(n_x+1, n_y, n_z|\mathbf{n}))\\
   \\
\dfrac{V_1^2}{V^2}(T(n_x, n_y-1, n_z|\mathbf{n})+T(n_x, n_y+1, n_z|\mathbf{n})) \\
\\
T(n_x, n_y, n_z-1|\mathbf{n})+T(n_x, n_y, n_z+1|\mathbf{n})
\end{pmatrix}
\end{equation}
Equation \eqref{eq:Fokker-Planck} admits the following equivalent Langevin representation
 \begin{equation} \label{eq:LangevinBcorr}
 \begin{cases}
     \dfrac{dx}{d\tau_1}=\dfrac{V_1}{V}(T(n_x+1, n_y, n_z|\mathbf{n})-T(n_x-1, n_y, n_z|\mathbf{n}))+\dfrac{1}{\sqrt{V_1}}\rho^{(1)}\\
     \\
     \dfrac{dy}{d\tau_1}=\dfrac{V_1}{V}(T(n_x, n_y+1, n_z|\mathbf{n})-T(n_x, n_y-1, n_z|, \mathbf{n}))+\dfrac{1}{\sqrt{V_1}}\rho^{(2)}\\
     \\
     \dfrac{dz}{d\tau_1}=T(n_x, n_y, n_z+1|\mathbf{n})-T(n_x, n_y, n_z-1|\mathbf{n})+\dfrac{1}{\sqrt{V_1}}\rho^{(3)}
     \end{cases}
 \end{equation}
 where $\boldsymbol{\rho}=(\rho^{(1)}, \rho^{(2)}, \rho^{(3)})$ is a stochastic vector whose components have zero mean and correlator $<\rho^{(\ell)}(\tau_1)\rho^{m}(\tau_1')>=B_\ell\delta_{\ell m}\delta(\tau_1-\tau_1')$. It is convenient to express the noise in terms of Gaussian white noise with unit variance. To this end, we introduce  $\boldsymbol{\lambda}=(\lambda^{(1)}, \lambda^{(2)}, \lambda^{(3)})$, satisfying
 \begin{eqnarray*}
     <\lambda^{(\ell)}> &= & 0\\
      <\lambda^{(\ell)}(\tau_1)\: \lambda^{(m)}(\tau_1')>&=&\delta_{\ell m}\delta(\tau_1-\tau_1') 
      \quad .
 \end{eqnarray*}
 Denoting by  $\mathbf{\mathcal{B}}$ the diagonal diffusion matrix, with entries  $\mathcal{B}_{ij}=B_i\delta_{ij}$ with $\mathbf{B}$ given by Eq. \eqref{eq:vectorB}, we perform the change of variables  $\boldsymbol{\rho}=G\boldsymbol{\lambda}$ where the  matrix $G$ is defined such  that $\mathbf{\mathcal{B}}=GG^t$. In this case $G$ is diagonal with entries
 \begin{equation*}
     \mbox{diag}G=
     \begin{pmatrix}
         \displaystyle{\frac{V_1}{V}\sqrt{T(n_x-1,n_y,n_z|\mathbf{n})+T(n_x+1, n_y, n_z|\mathbf{n})} }\\
         \\
          \displaystyle{\frac{V_1}{V}\sqrt{T(n_x, n_y-1, n_z|\mathbf{n})+T(n_x, n_y+1, n_z|\mathbf{n})}}  \\
          \\
        \displaystyle{ \sqrt{T(n_x, n_y,  n_z-1|\mathbf{n})+T(n_x, n_y, n_z+1|\mathbf{n})}}
     \end{pmatrix}
     \quad .
 \end{equation*}
 With this transformation, the Langevin equation \eqref{eq:LangevinBcorr} can be recast in the form of Eq. \eqref{eq:langevinKM}. 
\section{Linear noise approximation}\label{sec:LN}
We now expand the nonlinear terms in Eq.~\eqref{eq:langevinKM} in powers of $V_1^{-1/2}$. The arguments of the sigmoid functions become 
\begin{equation*}
    \begin{aligned}
        s_x= -r\dfrac{\xi_y}{\sqrt{V_1}} \\
        s_y=r\dfrac{\xi_x}{\sqrt{V_1}}
    \end{aligned}
\end{equation*}
which yield, at first order,
\begin{equation*}
    \begin{aligned}
    f(s_x)\approx f(0)+f'(0)\biggl(-r\dfrac{\xi_y}{\sqrt{V_1}}\biggr)= \dfrac{1}{1+b}-r\dfrac{b}{(1+b)^2}\dfrac{\xi_y}{\sqrt{V_1}}\\
    f(s_y)\approx f(0)+f'(0)\biggl(r\dfrac{\xi_x}{\sqrt{V_1}}\biggr)=\dfrac{1}{1+b}+r\dfrac{b}{(1+b)^2}\dfrac{\xi_x}{\sqrt{V_1}}
\end{aligned} \quad .
\end{equation*}
Neglecting higher-order terms in $V_1^{-1/2}$, Eq.~\eqref{eq:langevinKM} reduces to
\begin{align*}
    \sqrt{x+\gamma xz+f(s_x)}\approx \sqrt{\frac{1}{2}+\frac{\gamma\alpha_z}{2\delta_z}+\frac{1}{1+b}}=\sqrt{1+\gamma\frac{\alpha_z}{\delta_z}}, \\
    \\
    \sqrt{y+\gamma yz+f(s_y)}\approx \sqrt{\frac{1}{2}+\frac{\gamma\alpha_z}{2\delta_z}+\frac{1}{1+b}}=\sqrt{1+\gamma\frac{\alpha_z}{\delta_z}}, \\
    \\
   \sqrt{z\delta_z+\alpha_z}\approx \sqrt{2\alpha_z}. 
\end{align*}
In this way, neglecting higher order terms with respect to  $\displaystyle{\frac{1}{\sqrt{V_1}}}$, Eq. \eqref{eq:langevinKM} becomes  
\begin{equation}
\begin{cases}
    \dfrac{d\xi_x}{d\tau_1}=\dfrac{V_1}{V}\left [ \xi_x\biggl(-1-\gamma\dfrac{\alpha_z}{\delta_z}\biggr)-\xi_y\dfrac{r}{4}\biggl(1-\gamma^2\dfrac{\alpha_z^2}{\delta_z^2}\biggr)-\dfrac{\gamma}{2}\xi_z \right ]+\dfrac{V_1}{V}\sqrt{1+\gamma\dfrac{\alpha_z}{\delta_z}}\ \lambda^{(1)} \\
    \\
   \dfrac{d\xi_y}{d\tau_1}=\dfrac{V_1}{V} \left [ \xi_x\dfrac{r}{4}\biggl(1-\gamma^2\dfrac{\alpha_z^2}{\delta_z^2}\biggr)+\xi_y\biggl(-1-\gamma\dfrac{\alpha_z}{\delta_z}\biggr)-\dfrac{\gamma}{2}\xi_z \right ] +\dfrac{V_1}{V}\sqrt{1+\gamma\dfrac{\alpha_z}{\delta_z}} \ \lambda^{(2)}\\
    \\
    \dfrac{d\xi_z}{d\tau_1}=-\delta_z\xi_z+\sqrt{2\alpha_z} \ \lambda^{(3)}
    \end{cases}
\end{equation}
that corresponds to the extended version of Eq. \eqref{eq:Lang_lin}.

\begin{acknowledgments}
Francesca Di Patti thanks Gruppo Nazionale di Fisica Matematica (GNFM) of Istituto Nazionale di Alta Matematica (INDAM) for partial financial support. The work by Duccio Fanelli is supported by NEXTGENERATIONEU (NGEU) and funded by the Ministry of University and Research (MUR), National Recovery and Resilience Plan (NRRP), project MNESYS (PE0000006) ``A Multiscale integrated approach to the study of the nervous system in health and disease'' (DR. 1553 11.10.2022).
\end{acknowledgments}




\bibliographystyle{plain}
\bibliography{bibliography}

\end{document}